\documentclass[twocolumn,prb,showpacs,multicol,amsmath,amssymb]{revtex4}
\usepackage[dvips]{graphicx}
\usepackage{graphicx}
\usepackage{dcolumn}
\usepackage{bm}
\usepackage{graphics}
\usepackage{epsfig,color}

\newcommand{\be}{\begin{equation}}
\newcommand{\ee}{\end{equation}}
\newcommand{\bea}{\begin{eqnarray}}
\newcommand{\eea}{\end{eqnarray}}

\begin{document}

\title{The quantum compass chain in a transverse magnetic field}
\author{Mostafa Motamedifar, Saeed Mahdavifar, Saber Farjami Shayesteh}
\affiliation{ Department of Physics, University of
Guilan, 41335-1914, Rasht, Iran\\
}
\date{\today}

\begin{abstract}
We study the magnetic behaviors of a spin-1/2 quantum compass
chain (QCC) in a transverse magnetic field, by means of the
analytical spinless fermion approach and numerical Lanczos method.
In the absence of the magnetic field, the phase diagram is divided
into four gapped regions. To determine what happens by applying a
transverse magnetic field, using the spinless fermion approach,
critical fields are obtained as a function of exchanges. Our
analytical results show, the field-induced effects depend on in
which one of the four regions the system is. In two regions of the
phase diagram, the Ising-type phase transition happens in a finite
field. In another region, we have identified two quantum phase
transitions in the ground state magnetic phase diagram. These
quantum phase transitions belong to the universality class of the
commensurate-incommensurate phase transition. We also present a
detailed numerical analysis of the low energy spectrum and the
ground state magnetic phase diagram. In particular, we show that
the intermediate state ($h_{c_{1}}<h<h_{c_{2}}$) is gapful,
describing the spin-flop phase.
\\
\\
Email: m.motamedifa@guilan.ac.ir\\
$~~~~~~~~~~$ mahdavifar@guilan.ac.ir\\
$~~~~~~~~~~$       saber@guilan.ac.ir
\end{abstract}


\maketitle

\section{Introduction}\label{sec1}

Finding out the reason of different behaviors of quantum magnets
has attracted much interest in recent years. It is known that the
interaction between atoms plays a major role. Recently,
theoretical works are focused on the important role played by the
orbital degree of freedom in determining the magnetic properties
of quantum magnets. The complex interplay among the spin and
these types of degrees of freedom in quantum magnets, makes their phase
diagram rich and induces various fascinating physical phenomena. Also,
in some phenomena such as high-temperature superconductivity and
colossal magnetic resistance, the orbital degree of freedom of
electrons displays an important role.

The experimental observations on Mott insulators are a realization
of the effect of the orbital degree of freedom on the low-energy
behavior of a system. A proposed tactic to investigate these
systems is based on the so-called quantum compass
model\cite{Kugel73}. In fact, the quantum compass model is a very
good candidate for explaining the low-temperature behavior of some
Mott insulators. In this model, the orbital degrees of freedom are
represented by (pseudo)spin-1/2 operators and coupled
anisotropically in such a way to mimic the competition between
orbital ordering in different directions.

The two-dimensional quantum compass model is introduced as a
realistic model to generate protected cubits\cite{Doucot05} and
can play a role in the quantum information theory. This model is
dual to studied model of superconducting arrays\cite{Nussinov05}.
It was shown that the eigenstates are two-fold degenerate and to
be gapped\cite{Doucot05}. The results from both spin wave study
and exact diagonalization have suggested a first-order quantum
phase transition in the ground state phase diagram\cite{Dorier05}.
The existing of the first-order phase transition is confirmed in
two-dimensional quantum compass model\cite{Chen07,Otus09}.
Recently, it is found that in the ground state, most of the
two-site spin correlations vanish and the two-dimer correlations
exhibit the nontrivial hidden order\cite{Brzezicki10}. Another
examination of the 2D quantum compass model showed that a second
order phase transition can occur when the frustration of exchange
interaction increases\cite{Cincio10}. In addition, finite
temperature properties have been studied. It is found that the
low-temperature ordered phase with a thermal transition
corresponds to the 2D Ising universality class\cite{Wenzel08,
Mishra04} and due to the dilution, the decrease of ordering
temperature is much stronger than that in spin
models\cite{Tanaka07}.

The 1D version is known as the quantum compass chain (QCC) and
represents one particular subclass of low-dimensional quantum
magnets which poses interesting theoretical
problems\cite{Brzezicki07,You08,Brzezicki09,Sun2-09,Eriksson09,
Mahdavifar10}.
By mapping the model to a quantum Ising chain, an exact solution
is obtained for the ground state energy and the complete
excitation spectrum\cite{Brzezicki07}. It is shown that the QCC
exhibits a first-order phase transition at $J_1=0$ between two
disordered phases with opposite signs of certain local spin
correlations. The model is also diagonalized exactly by a direct
Jordan-Wigner transformation\cite{Brzezicki09}. The obtained
results by latter approach, confirm the existence of the
first-order phase transition in the ground state phase diagram. In
a very interesting work, it is found that the reported first-order
phase transition, in fact occurs at a multicritical point where a
line of the first-order transition meets with a line of the
second-order transition\cite{Eriksson09}.   Based on a numerical
analysis\cite{Mahdavifar10}, the first and second order quantum
phase transitions in the ground state phase diagram have been
identified. By a detailed analysis of the numerical results on the
spin structure factors, it is shown that the N$\acute{e}$el and
the stripe-antiferromagnet long-range orders exist in the ground
state phase diagram\cite{Mahdavifar10}.

The effect of a transverse magnetic field on the QCC is studied by
Ke-Wei Sun et al.\cite{Sun1-09}. They allowed the changing a
parameter that only provides moving on the
 first order critical line. Opening the energy gap of the system
when a magnetic field applied is the result that they could
achieve. In spite of the calculation of pseudo-spin correlation
functions, fidelity susceptibility, the concurrence, the
block-block entanglement entropy, they didn't investigate the
ground state phase diagram of the QCC in a transverse magnetic
field. They only restricted their study on the first order
critical line. In a very recent work\cite{Motamedifar10}, our
group studied  the presence QCC in an area of the ground state
magnetic phase diagram where the odd couplings are
antiferromagnetic and larger than even couplings. Using pseudo
spin ladder operators indicated that the QCC in a transverse
magnetic field can be mapped to the one dimensional Ising model in
a transverse magnetic field (ITF). Analytical investigation of the
effective ITF Hamiltonian predicts the occurrence of two quantum
phase transitions by increasing magnetic field.

In this paper we use the main idea of Perk\cite{Perk75}  to
provide analytical results on the effect of a transverse magnetic
field in whole range of the ground state phase diagram. Depending
on values of the couplings, one or two quantum phase transitions
can happen in finite critical fields. To make a clear picture of
different long-range order phases, we do a very accurate numerical
experiment. In particular we use the numerical Lanczos method to
diagonalize numerically finite chains up to $N=24$ spins. Based on
the exact diagonalization results we show that the spin-flop
long-range order induces between two critical magnetic fields. The
spin-flop phase is denoted as a phase with the staggered
magnetization perpendicular to the direction of the transverse
magnetic field.

The outline of the paper is as follows. In forthcoming section
 we introduce the results of the analytical fermionization studies.
 In the section III we discuss the model in the
strong exchange couplings limit and derive the effective spin
chain Hamiltonian. In section IV, we present results of an
accurate numerical experiment. Finally, we have discussed and
summarized our results in section V.



\section{Fermionization}\label{sec2}

The Hamiltonian of the QCC in a transverse magnetic field with $N$
periodic sites is given by\cite{Eriksson09, Mahdavifar10}
\begin{eqnarray}
{\cal H}&=&\sum_{i=1}^{N/2} J_{1} \sigma_{2i-1}^{z}\sigma_{2i}^{z}
+ J_{2}\sigma_{2i-1}^{x} \sigma_{2i}^{x}+ L \sigma_{2i}^{z}
\sigma_{2i+1}^{z}\nonumber \\
&-&h \sum_{i=1}^{N}\sigma_{i}^{y}. \label{Hamiltonian}
\end{eqnarray}
Here $\sigma_{i}^{x,z}$ are the Pauli operators on the $i$th site
and $J_1, J_2, L_1$ are the exchange couplings. $h$ denotes the
transverse magnetic field. In the absence of the transverse
magnetic field, $h=0$, the first and second order critical lines
denote with $J_1/L_1=0$ and $J_2/L_1=1$,
respectively\cite{Eriksson09}. There are four gapped phases in the
regions: (I.)$J_1/L_1>0$, $J_2/L_1>1$, (II.) $J_1/L_1>0$,
$J_2/L_1<1$, (III.) $J_1/L_1<0$, $J_2/L_1<1$, (IV.) $J_1/L_1<0$,
$J_2/L_1>1$. There is a hidden order in the regions (I.) and
(IV.). But two kind of magnetic long-range orders, a type of the
N$\acute{e}$el and the stripe-antiferromagnetic orders have been
recognized in the regions (II.) and (III.)
respectively\cite{Mahdavifar10}.

A mapping of the spin-$\frac{1}{2}$ operators onto fermi operators
by means of the jordan-Wigner transformation was vastly used by
many physicists who are interested to investigate the Hamiltonian
of the system by the aid of the spinless fermions. Because there
are alternating links in the QCC we introduce two kinds of
spinless fermion through the following Jordan-Wigner
transitions\cite{Abouie08} :
\begin{eqnarray}
{\cal S}_{2n-1}^{+}&=&a^{\dagger}e^{i\pi\sum_{m=1}^{n-1} (a^{\dagger}_{m}a_{m}+ b^{\dagger}_{m}b_{m})}\nonumber \\
{\cal S}_{2n}^{+}&=&b^{\dagger}e^{i\pi(\sum_{m=1}^{n} a^{\dagger}_{m}a_{m}+ \sum_{m=1}^{n-1}b^{\dagger}_{m}b_{m})}\nonumber \\
{\cal S}_{2n-1}^{y}&=&a^{\dagger}_{n}a_{n}-\frac{1}{2},{\cal
S}_{2n}^{y}=b^{\dagger}_{n}b_{n}-\frac{1}{2}.
\label{fermionization}
\end{eqnarray}
Using the above transformation, the QCC is mapped to a 1D
interacting spinless fermion system:

\begin{eqnarray}
{\cal H}_{f}&=&(\frac{J_{1}-J_{2}}{4})\sum_{n=1}^{\frac{N}{2}}[a^{\dagger}_{n}b^{\dagger}_{n}-a_{n}b_{n}]\nonumber\\
&+&(\frac{J_{1}+J_{2}}{4})\sum_{n=1}^{\frac{N}{2}}[ a^{\dagger}_{n}b_{n}-a_{n}b^{\dagger}_{n}]\nonumber \\
&+&\frac{L}{4}\sum_{n=1}^{\frac{N}{2}}[b^{\dagger}_{n}a^{\dagger}_{n+1}-b_{n}a_{n+1}-b_{n}a^{\dagger}_{n+1}+b^{\dagger}_{n}a_{n+1}]\nonumber \\
&-&h\sum_{n=1}^{\frac{N}{2}}[(a^{\dagger}_{n}a_{n}-\frac{1}{2})+(b^{\dagger}_{n}b_{n}-\frac{1}{2})].
\label{fermionization2}
\end{eqnarray}
Then by means of Fourier transformations, the mean field
Hamiltonian is given by

\begin{eqnarray}
{\cal H}_{MF}&=&(\frac{J_{1}-J_{2}}{4})\sum_{q}[A^{\dagger}_{q}B^{\dagger}_{-q}+B_{-q}A_{q}]\nonumber\\
&+&\sum_{q}(\frac{J_{1}+J_{2}+Le^{-iq}}{4})[A^{\dagger}_{q}B_{q}]\nonumber\\
&+&\sum_{q}(\frac{J_{1}+J_{2}+Le^{+iq}}{4})[B^{\dagger}_{q}A_{q}]\nonumber\\
&+&(\frac{L}{4})\sum_{q}[e^{+iq}B^{\dagger}_{q}A^{\dagger}_{-q}+e^{-iq}A_{-q}B_{q}]\nonumber\\
&-&h\sum(A^{\dagger}_{q}A_{q}+B^{\dagger}_{q}B_{q}).
\label{fermionization2}
\end{eqnarray}

Finally, by diagonalizing the mean field Hamiltonian\cite{Perk75},
the critical transverse fields obtain as

\begin{eqnarray}
h_{c_{1}}&=&\sqrt{J_{1}(L+J_{2})}\nonumber\\
\nonumber\\
h_{c_{2}}&=&\sqrt{-J_{1}(L-J_{2})}.\label{critical-fields}
\end{eqnarray}
Surprizing is that, by applying the transverse magnetic field two
quantum phase transitions can occur in the region (I.). But in the
regions (II.) and (III.), there is only one critical field. In the
region (IV.), no quantum phase transition occurs by increasing the
transverse magnetic field in a finite value of $h$.

\begin{figure*}[t]
\centerline{\includegraphics[width=16cm,height=16cm,angle=0]{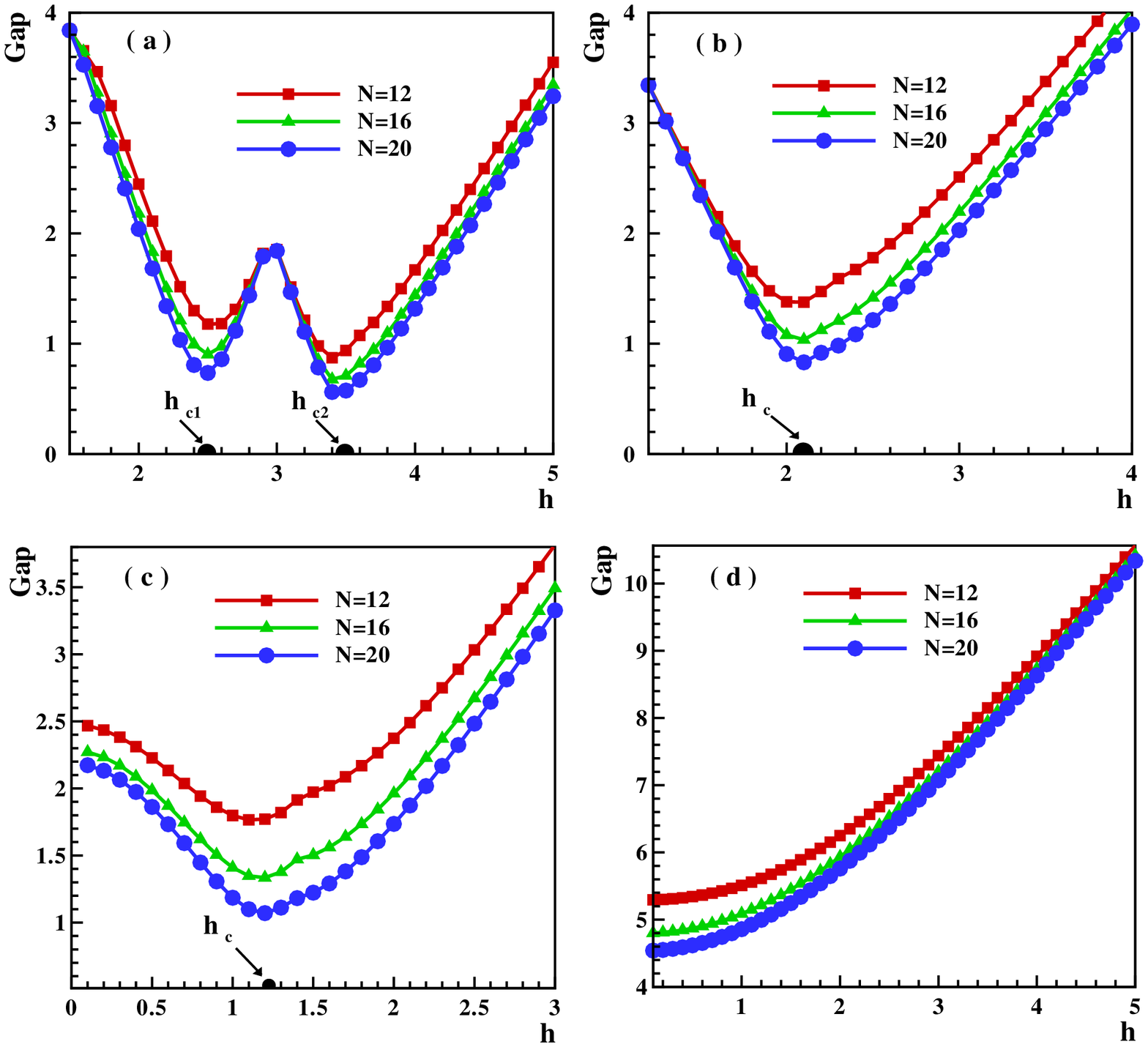}
}\caption{(Color online.) The energy gap as a function of the
transverse magnetic filed $h$, for different chain lengths $N=12,
16, 20$ and exchanges $L_{1}=1.0$, (a)$J_1=J_2=3.0$,
(b)$J_{1}=3.0, J_2=0.5$, (c) $J_{1}=-3.0, J_2=0.5$ and (d)
$J_{1}=-3.0, J_2=3.0$ .} \label{E-Gap}
\end{figure*}

\section{Effective Hamiltonian} \label{sec2}

In this section, we briefly discuss the model (\ref{Hamiltonian})
in limiting case of the exchanges
$\frac{J_1}{L}=\frac{J_2}{L}=\frac{J}{L}\gg 0$. In this limit the
model (\ref{Hamiltonian}) can be mapped onto the effective spin
chain Hamiltonian\cite{Motamedifar10}.

In order to work easily, we rotated the system in the special sort
that the field oriented in $z$  direction. In the absence of the
magnetic field, we consider one of the odd links for example the
first link. For this dimer the quantum state is either the singlet
state, $|S\rangle$,  with the energy eigenvalue $-J/2$  or triplet
state, $|T_{1}\rangle$, $|T_{-1}\rangle$, $|T_{0}\rangle$ with
energy eigenvalues of $0$, $0$, $J/2$ respectively.

At $h=0$  the ground state of a distinct dimer is $|S\rangle$. As
the magnetic field $h$ increases the energy of the triplet state
$|T_1\rangle$ decreases and at $h=J/2$ forms together with the
singlet state, a doublet of almost degenerate low energy state.
This degenerate state splits from the remaining high energy two
triplet states. Thus, for a strong enough magnetic field we have a
situation when the singlet $|S\rangle$  and triplet $|T_1\rangle$
states create a new effective spin $\tau=1/2$ system. On the new
singlet-triplet subspace and up to a constant, one can easily
obtain the effective Hamiltonian\cite{Motamedifar10}

\begin{eqnarray}
{\cal H}^{eff}=\frac{L}{2} \sum_{n=1} \tau_{n}^{x}
\tau_{n+1}^{x}-(h-J)\sum_{n=1}\tau_{n}^{z}.
\label{Eff-Hamiltonian}
\end{eqnarray}

The effective Hamiltonian describes the Ising chain in an
effective transverse magnetic field\cite{Vojta03}$h^{eff}=h-J$.
The critical effective fields for the Ising chain in a transverse
magnetic field (ITF) are rigorously interesting, one of them is
$h_{c_{1}}^{eff}=-L/2$ and the other one is $h_{c_{2}}^{eff}=L/2$.
Therefore critical fields of the main QCC can be obtained as

\begin{eqnarray}
h_{c_{1}}&=&J-L/2\nonumber\\
\nonumber\\
h_{c_{2}}&=&J+L/2.\label{Eff-critical-fields}
\end{eqnarray}

It is clear that the hidden ordered gapped phase at
$h^{eff}<h^{eff}_{c_{1}}$ for the QCC corresponds to the
negatively saturated magnetization phase for the effective ITF
model. Whereas the intermediate gapped phase at
$h^{eff}_{c_{1}}<h^{eff}<h^{eff}_{c_{2}}$ for the QCC corresponds
to the N$\acute{e}$el phase for the effective ITF model. Finally,
the region $h^{eff}>h^{eff}_{c_{2}}$ corresponds to the fully
magnetization phase of the effective ITF model where the QCC is
totally magnetized. In the next section, we present our numerical
results obtained by a very accurate numerical experiment and
confirm the mentioned quantum phase transitions.

\section{Numerical experiment} \label{sec3}

By doing an experiment, the validity of the suggested theoretical
results on the effect of a transverse magnetic field on  the QCC,
can be determined. Since a real experiment cannot be done at zero
temperature, the best way is doing a virtual numerical experiment.
A very famous and accurate method in field of the numerical
experiments is known as the Lanczos method. The Lanczos method and
the related recursion methods\cite{Lanczos50,Grosso95}, possibly
with appropriate implementations, have emerged as one of the most
important computational procedures, mainly when a few extreme
eigenvalues are desired. However, the strong role of a numerical
experiment to examine quantum phase transitions is not negligible.

In this section, to explore the nature of the spectrum and the
quantum phase transition, we used Lanczos method to diagonalize
numerically chains with length up to $N=24$ and different values
of the exchanges. The energies of the few lowest eigenstates were
obtained for chains with periodic boundary conditions.

The first information that can be gained is the energy gap.  In
this way we have accessed to the energy gap which is recognized as
the difference between the second exited state and the ground
state energy in finite QCC. In Fig.\ref{E-Gap} the energy gap is
plotted versus transverse magnetic field, $h$, for various length
of chains. This figure contains four graphs of the energy gap in
different sectors of the ground state phase diagram. The
Fig.\ref{E-Gap}(a) devoted to the energy gap of the system at the
area (I.) of the phase diagram. In this figure the computed energy
gap corresponding various length of chains ($N=12, 16, 20$) and
exchanges $L=1.0$, $J_1=3.0$ and $J_2=3.0$  is plotted versus the
transverse magnetic field. By increasing the size of the system
the energy gap curve will approximate to the magnetic field axes.
The magnetic fields in which the energy gap in the thermodynamic
limit will be closed, are critical fields. By means of the
phenomenological renormalization group
technique\cite{Mahdavifar08-1} we have found $h_{c1}=2.5\pm0.1$
and $h_{c2}=3.5\pm 0.1$ for critical fields in that the system
undergos the quantum phase transition. In other words, in absence
of the transverse magnetic field, the system is in a gapfull
phase. By increasing the magnetic field, the gap of energy
decreases and finally will close at the first critical field
$h_{c1}=2.5\pm0.1$. After first critical field the gap of energy
will be opened by growing magnetic field, while at the second
critical field $h=3.5\pm0.1$, again will be closed.  For values of
the magnetic field further than $h=3.5\pm0.1$ the gap of system
will open and the system will go to a saturated phase. In
conclusion, applying the transverse magnetic field to the gapped
QCC in region (I.), creates  two new gapfull magnetic phases. The
numerical critical fields with selected amount of exchanges $L$,
$J_{1}$ and $J_{2}$ precisely correspond with the previous
analytical critical fields (Eq.\ref{critical-fields} and
Eq.\ref{Eff-critical-fields}).

\begin{figure*}[t]
\centerline{\includegraphics[width=16cm,height=16cm,angle=0]{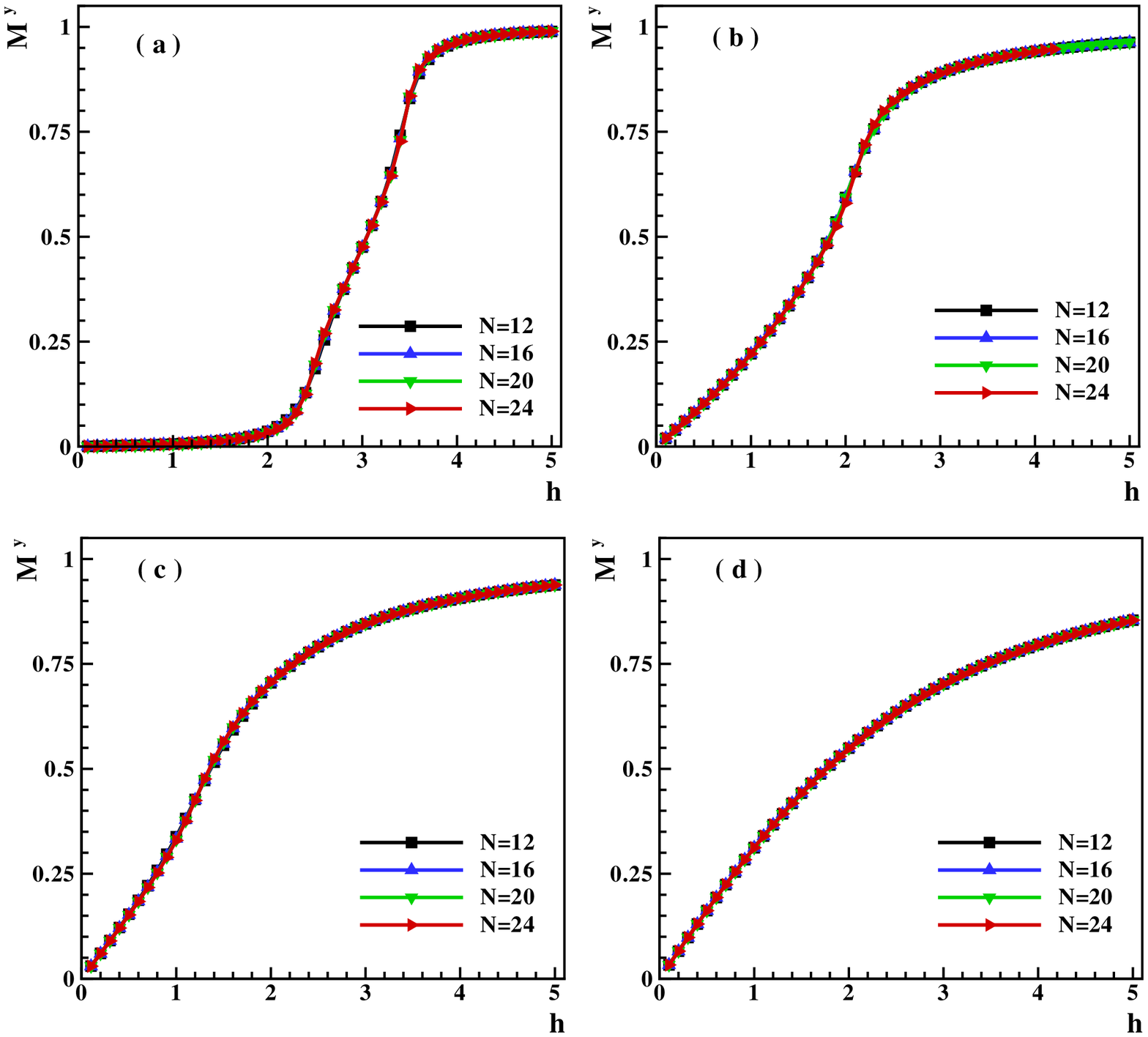}
}\caption{(Color online.) The magnetization along the field
$M^{y}$, as a function of the transverse magnetic filed $h$, for
different chain lengths $N=12, 16, 20, 24$ and exchanges
$L_{1}=1.0$, (a)$J_1=J_2=3.0$, (b)$J_{1}=3.0, J_2=0.5$, (c)
$J_{1}=-3.0, J_2=0.5$ and (d) $J_{1}=-3.0, J_2=3.0$. }
\label{Magnetization}
\end{figure*}

In figures \ref{E-Gap}(b) and \ref{E-Gap}(c),  the numerical
results on the energy gap are plotted for various length of
chains, $N=12, 16, 20$ and exchanges $L=1.0$, $J_1=3.0$,
$J_2=0.5$, in region (II.) and $J_1=-3.0$, $J_2=0.5$, in region
(III.). As it can be seen in Fig.\ref{E-Gap}(b) and (c), the
qualitative effect of the transverse magnetic field  in the
regions (II.) and (III.) is equivalent. In the absence of the
magnetic field, the system is gapped. Increasing the transverse
magnetic filed, the energy gap decreases and will be closed in
$h_{c}=2.1\pm0.1$ ($h_{c}=1.2\pm0.1$) in regions (II.) and (III.)
respectively.  More increasing the magnetic field from $h_c$, the
energy gap opens again and behaves almost linearly. The obtained
numerical critical fields in these regions are in well agreement
with the analytical results (Eq.\ref{critical-fields}). Finally,
Fig.\ref{E-Gap}(d) allocated to the energy gap at the area (IV.)
of the ground state phase diagram for chain sizes $N=12, 16, 20$,
and exchanges $L=1.0$, $J_1=-3.0$, $J_2=3.0$. In this figure, is
clearly seen that by increasing the transverse magnetic field, the
energy spectrum of the QCC will remain gapfull. There aren't any
critical magnetic fields. The system goes to a saturated magnetic
phase for enough large values of the transverse magnetic field.

So far the paper was determined that the transverse field,
depending on the values of the exchanges can induces two or one
gapped phases in the ground state phase diagram of the QCC.
Subsequently, a very important question that precedes is: "What
are the magnetic long-range ordered phases in these gapped regions
of the QCC?" A very clear answer can provides, doing the numerical
Lanczos experiment. To recognize the different magnetic phases
induced by transverse field, $h$, we implemented the algorithm for
finite-size chains ($N=12, 16, 20, 24$) to calculate the order
parameters and the various spin correlation functions.

An insight into the nature of different phases can be obtained by
studying the magnetization process. The magnetization along the
transverse field axis is defined as
\begin{eqnarray}
M^{y}= \frac{1}{N}\sum_{i=1}^{N}\langle Gs| \sigma_{i}^{y}
|Gs\rangle,\label{magnetization}
\end{eqnarray}
where the notation $\langle Gs|...|Gs\rangle$ represents the
ground state expectation value. In Fig.\ref{Magnetization}, we
have plotted the transverse magnetization, $M^{y}$, versus $h$,
for different chain sizes $N=12, 16, 20, 24$ and different values
of the exchanges that covered all regions in the ground state
phase diagram. We did not find any size effect on the numerical
results of the transverse magnetization. As is seen in
Fig.\ref{Magnetization}(a), due to the profound effect of quantum
fluctuations, the transverse magnetization remains small but
finite for $0<h<h_{c_{1}}$ and reaches zero at $h=0$. This
behavior shows that in the gapped phase with hidden order in
region (I.), the transverse magnetization appears only at a finite
critical value of the transverse field, $h_{c_{1}}$. This
phenomenon is also reported for a two-leg
ladder\cite{Mahdavifar07} and an alternating spin
chain\cite{Mahdavifar08-1}. For $h>h_{c_{1}}$, transverse
magnetization increases almost linearly with increasing transverse
field. At the second critical field $h_{c_{2}}$, due to the
quantum fluctuations the magnetization do not saturate and
reaching saturation asymptotically in the limit of infinite
transverse field.

Numerical results on the magnetization function presented in
figures \ref{Magnetization}(b) and \ref{Magnetization}(c) are
computed for various length of chains ($N=12, 16, 20, 24$) and
exchanges $L=1.0$, $J_1=3.0$, $J_2=0.5$, in region (II.) and
$J_1=-3.0$, $J_2=0.5$, in region (III.). As the same as the energy
gap function, the qualitative effect of the transverse magnetic
field in the regions (II.) and (III.) is equivalent. In the
absence of the transverse field, the transverse magnetization is
zero. As soon as the transverse field is applied, the transverse
magnetization start to increase. In the region (III.), the
transverse magnetization increases faster than the region(II.). It
is also completely clear that the system at the critical field
$h_c$ do not saturate. The saturation will be happened for very
larger values of the transverse field. Fig.\ref{Magnetization}(d)
shows the numerical results in the region (IV.) of the ground
state phase diagram for chain sizes $N=12, 16, 20, 24$, and
exchanges $L=1.0$, $J_1=-3.0$, $J_2=3.0$. As soon as the
transverse field is applied, the transverse magnetization starts
to increase from zero. The transverse magnetization grows smoothly
by increasing the transverse field and due to profound effect of
the quantum fluctuations, it will be saturated in infinite
transverse field. By comparing figures \ref{Magnetization}(a) and
\ref{Magnetization}(d), one can find that the hidden order in the
region (I.) must be completely different from the hidden order
phase in the region (IV.). The hidden order phase in the region
(I.), will not surrender versus the transverse field, but the
hidden order phase in the region (IV.) as soon as the transverse
field applied will surrender versus field and spins start to aline
the field direction.

To find the long-range magnetic order of the ground state of the
system, we start our consideration with the spin-spin correlation
function defined by
\begin{eqnarray}
W_{j}^{\alpha\alpha}(n) =
\langle\sigma_{j}^{\alpha}\sigma_{j+n}^{\alpha}\rangle~~~(\alpha=x,
y, z),
\end{eqnarray}
and the spin structure factor at momentum $q$ defined by
\begin{eqnarray}
S^{\alpha\alpha}(q) = \sum_{n=1}^{N-1}W_{j}^{\alpha\alpha}(n)
exp(iqn).
\end{eqnarray}
It is known that the spin structure factor give us a deep insight
into the characteristics of the ground state\cite{Mahdavifar10}.

\begin{figure}
    \centering
    {
        \includegraphics[width=8.0cm]{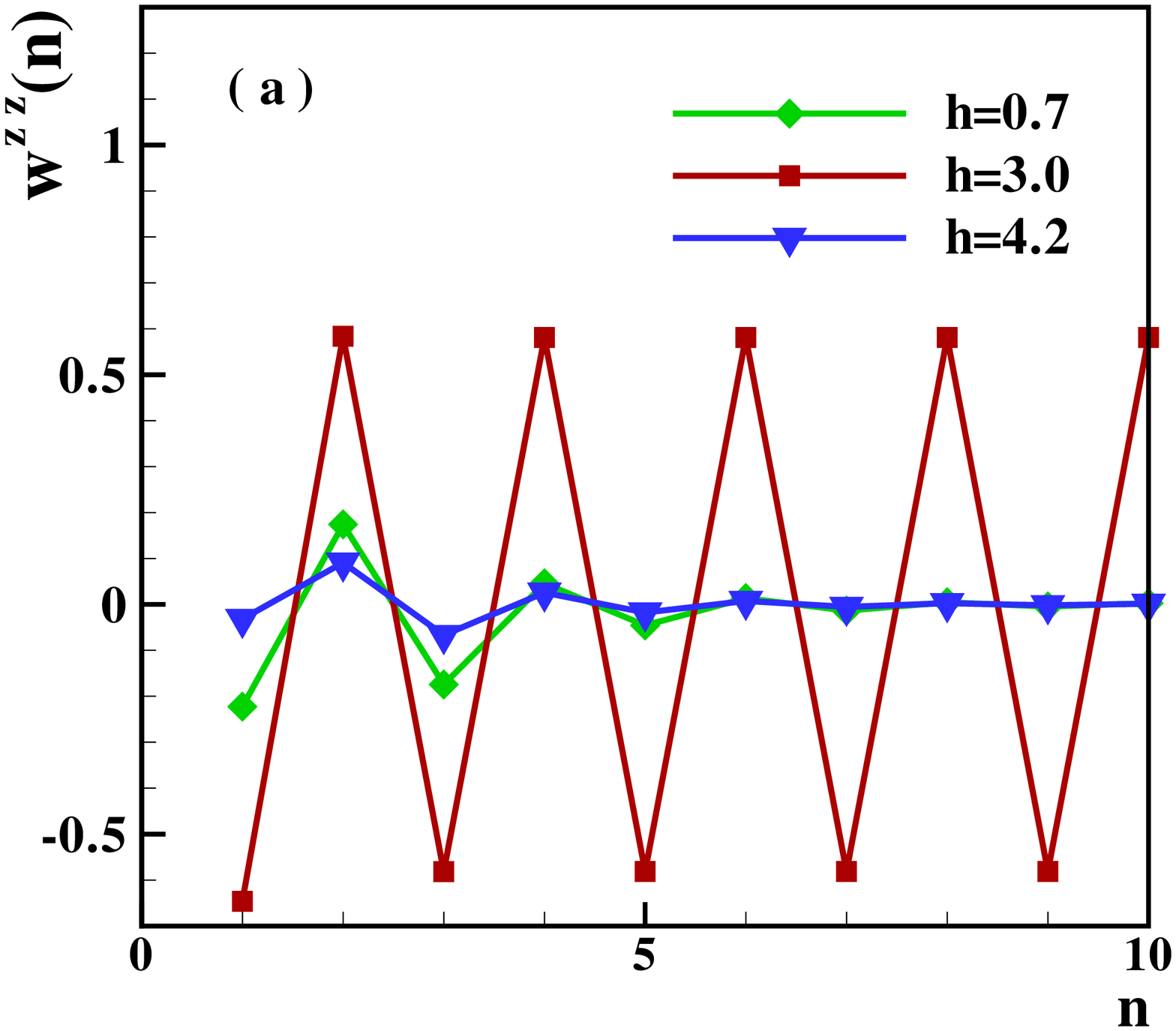}
        \label{fig:first_sub}
    }
    {
        \includegraphics[width=8.0cm]{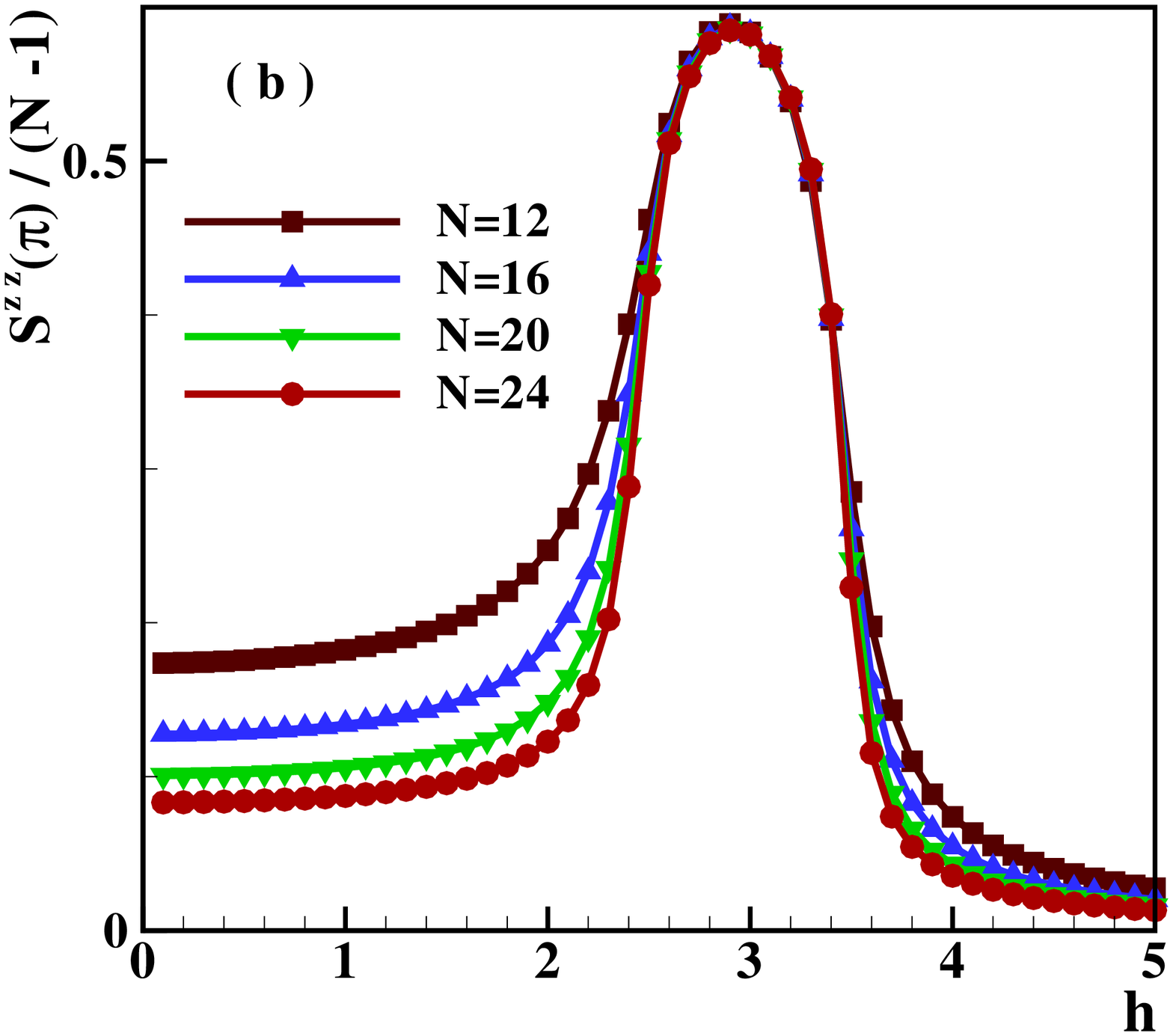}
        \label{fig:first_sub}
    }
    {
        \includegraphics[width=8.0cm]{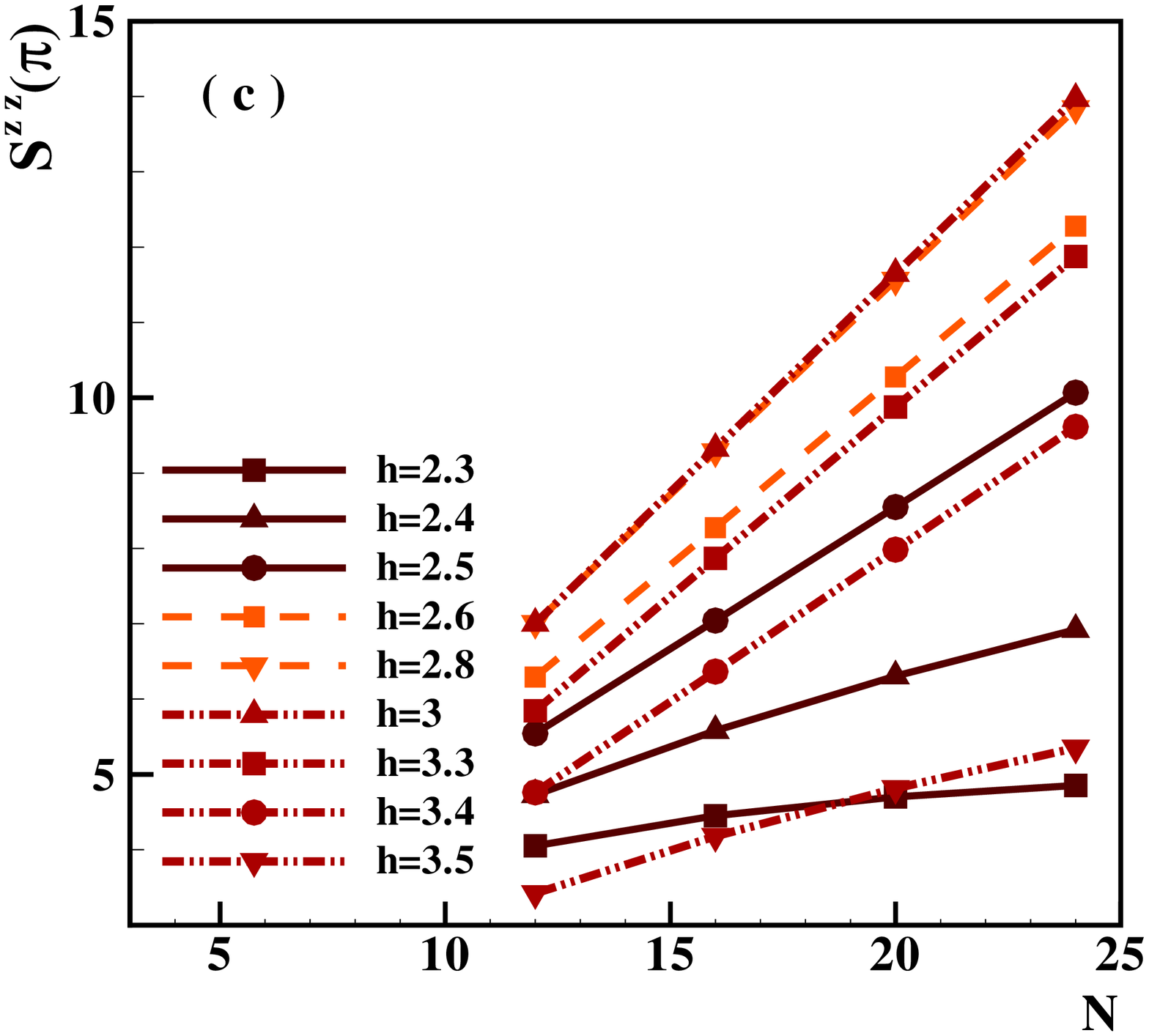}
        \label{fig:_sub}
    }
    \caption{(a) The spin-spin correlation $W_{1}^{zz}$ as a
    function of $n$, for chain length $N=20$ and exchanges $L_{1}=1.0$, $J_1=J_2=3.0$. (b) The spin structure factor
    $S^{zz}(\pi)/(N-1)$ as a function of the transverse field $h$ for different chain lengths
    $N=12, 16, 20, 24$ and exchanges $L_{1}=1.0$, $J_1=J_2=3.0$. (c) The spin structure factor $S^{zz}(\pi)$ plotted
    as a function of the chain length $N$, for different values of the field in the intermediate region.
    }
    \label{fig:Structure.neel.I}
\end{figure}

By examining the energy gap we found that a new magnetic phase is
induced by applying the transverse field in the region (I.)
between $h_{c_{1}}<h<h_{c_{2}}$. To determine how to long-range
order there is, in Fig.\ref{fig:Structure.neel.I}(a) we have
plotted $W_{1}^{zz}$ as a function of $n$ for different values of
the transverse field. The selected values of the transverse field
$h=0.7, 3.0, 4.2$ in the figure cover all phases in the region
(I.). It can be seen from this figure
(Fig.\ref{fig:Structure.neel.I}(a)) that the $z$ component of the
spins on odd sites  is pointed in the same direction with the
$\sigma_{1}^{z}$ and others (on even sites) are pointed in
opposite direction  for a value of $h=3.0$. This is an indication
for the N$\acute{e}$el ordering in the intermediate region
$h_{c_{1}}<h<h_{c_{2}}$. This is very interesting that the
numerical results suggest that a different novel phase can be
induced by increasing the transverse field which is known as the
spin-flop phase.
For any value of the transverse field the Lanczos results lead to
the staggered magnetization,
$M_{st}^{z}=\frac{1}{N}\sum_{j=1}^{N}(-1)^{j}\sigma_{j}^{z}=0$,
since the ground state is degenerate and in a finite system no
symmetry breaking happens. But, the $h$-dependency of the spin
structure factor, $S^{zz}(\pi)$, is qualitatively the same as the
staggered magnetization, $M_{st}^{z}$. However the spin-spin
correlation function diverges in the ordered phase as
$N\longrightarrow\infty$. We computed the spin structure factor
$S^{zz}(\pi)$ and in Fig.\ref{fig:Structure.neel.I}(b), we have
plotted $S^{zz}(q=\pi)/(N-1)$ as a function of $h$ for the chain
lengths $N=12, 16, 20, 24$ and exchanges $L=1.0$, $J_1=3.0$,
$J_2=3.0$. It can be seen that in the intermediate region
$h_{c_{1}}<h<h_{c_{2}}$, the ground state of the system is in the
N$\acute{e}$el phase. There is not the N$\acute{e}$el ordering
along $z$ axis in regions $h<h_{c_{1}}=2.5\pm0.1$ and
$h>h_{c_{2}}=3.5\pm0.1$. To check the existence of the
N$\acute{e}$el order in the thermodynamic limit $N \longrightarrow
\infty$ of the system in the intermediate region
$h_{c_{1}}<h<h_{c_{2}}$, we have plotted in
Fig.\ref{fig:Structure.neel.I}(c) the $N$ dependence of
$S^{zz}(\pi)$ for different values of transverse field. Increasing
behavior of the spin structure factor in the sector
$h_{c_{1}}<h<h_{c_{2}}$, shows that the spin-flop phase is a true
long-range order.

On the other hand the behavior of the energy gap versus the
magnetic field in the regions (II.) and (III.), suggested a
quantum phase transition at the critical field $h_{c}$. In a very
recent work\cite{Mahdavifar10}, N$\acute{e}$el and
stripe-antiferromagnetic long-range orders, in the absence of the
field, have been reported in the regions (II.) and (III.)
respectively. The "stripe-antiferromagnetic phase" is define as a
phase with the opposite magnetization along the $z$ axis on the
odd bonds. The order parameter of the stripe-antiferromagnetic
phase is defined as\cite{Mahdavifar10}
\begin{eqnarray}
M_{sp}^{z} = \frac{2}{N}\langle
\sum_{j=1}^{N/2}(-1)^{j}(\sigma_{2j-1}^{z}+ \sigma_{2j}^{z})
\rangle.
\end{eqnarray}
We computed the spin structure factor $S^{zz}(\pi)$ and the
correlation function of the stripe-antiferromagnetic order
parameter given by
\begin{eqnarray}
\chi^{zz} = \langle \sum_{n=1}^{N/2-1}(-1)^{n}(\sigma_{2j-1}^{z}+
\sigma_{2j}^{z}) (\sigma_{2j-1+2n}^{z}+ \sigma_{2j+2n}^{z})
\rangle. \nonumber \\
\end{eqnarray}
\begin{figure}[t]
\centerline{\psfig{file=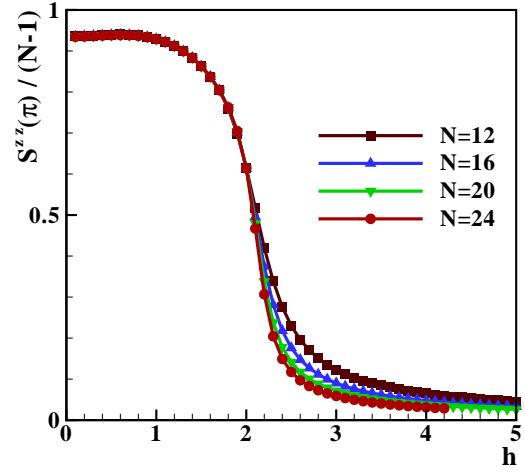,width=8.0cm}}
\caption{(Color online.)The spin structure factor
    $S^{zz}(\pi)/(N-1)$ as a function of the transverse field $h$ for different chain lengths
    $N=12, 16, 20, 24$ and exchanges $L_{1}=1.0$, $J_1=3.0, J_2=0.5$.} \label{Structure.neel.II}
\end{figure}

In Fig.\ref{Structure.neel.II}, we have plotted
$S^{zz}(\pi)/(N-1)$  as a function of $h$ for the chain lengths
$N=12, 16, 20, 24$ and exchanges $L=1.0$, $J_1=3.0$, $J_2=0.5$.
Clearly be seen that the N$\acute{e}$el order remains in the
presence of the field up to the critical field $h_c=2.1\pm0.1$.
Overlapping of the numerical results in the region
$h<h_c=2.0\pm0.1$ shows a divergent behavior of the function
$S^{zz}(\pi)$ by increasing the size of chain $N$. This justifies
that the N$\acute{e}$el ordering along the $z$ axis is true
long-range order in the region $h<h_c=2.1\pm0.1$ of the ground
state phase diagram. Induced quantum fluctuations by increasing
$h$ from zero, decreases the staggered magnetization from the
almost saturation value. We checked our numerical results and
found that there is not the N$\acute{e}$el ordering along $z$ axis
in the region $h>h_c$.

\begin{figure*}[t]
\centerline{
\includegraphics[width=8cm,height=8cm,angle=0]{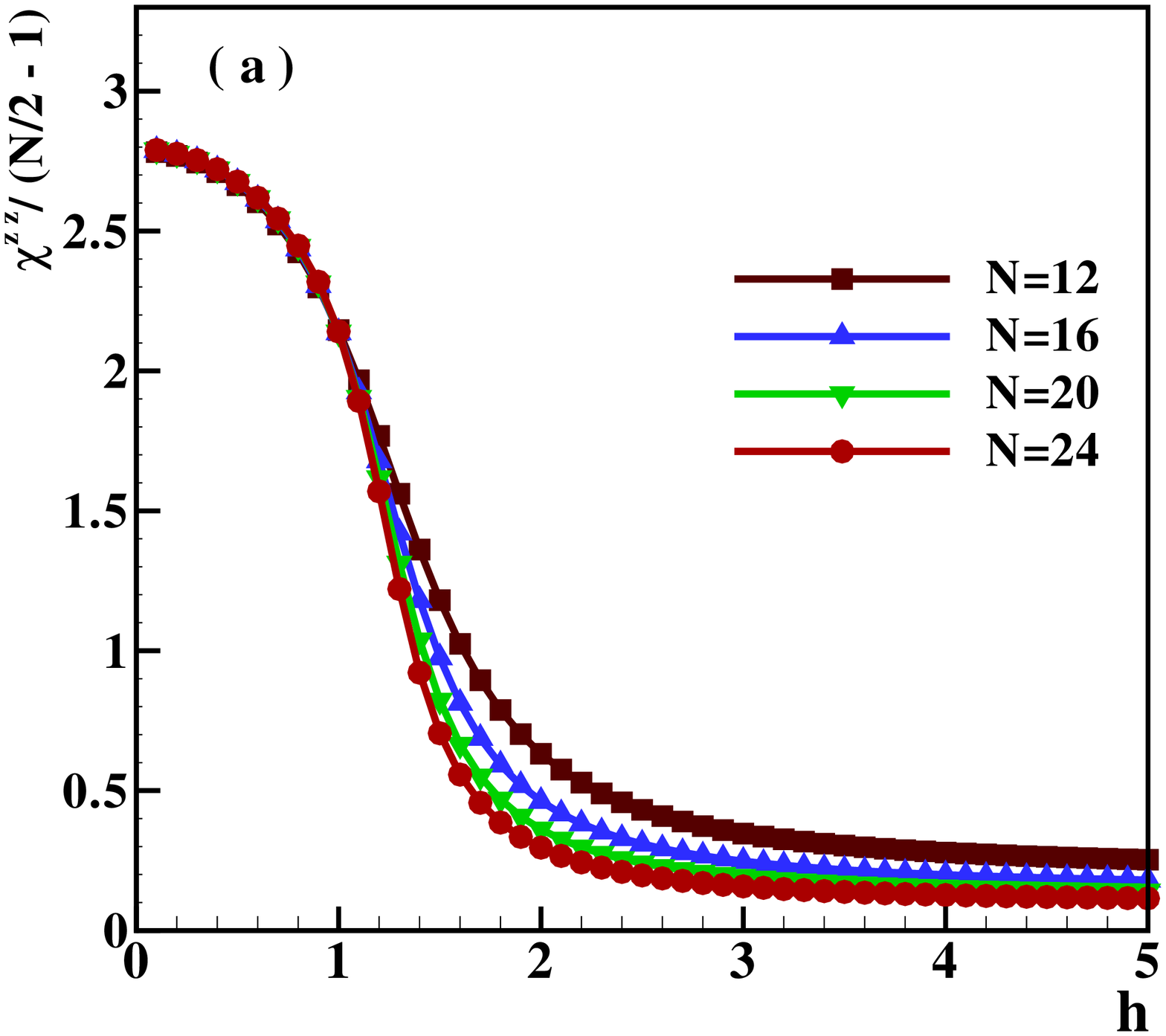}
\includegraphics[width=8cm,height=8cm,angle=0]{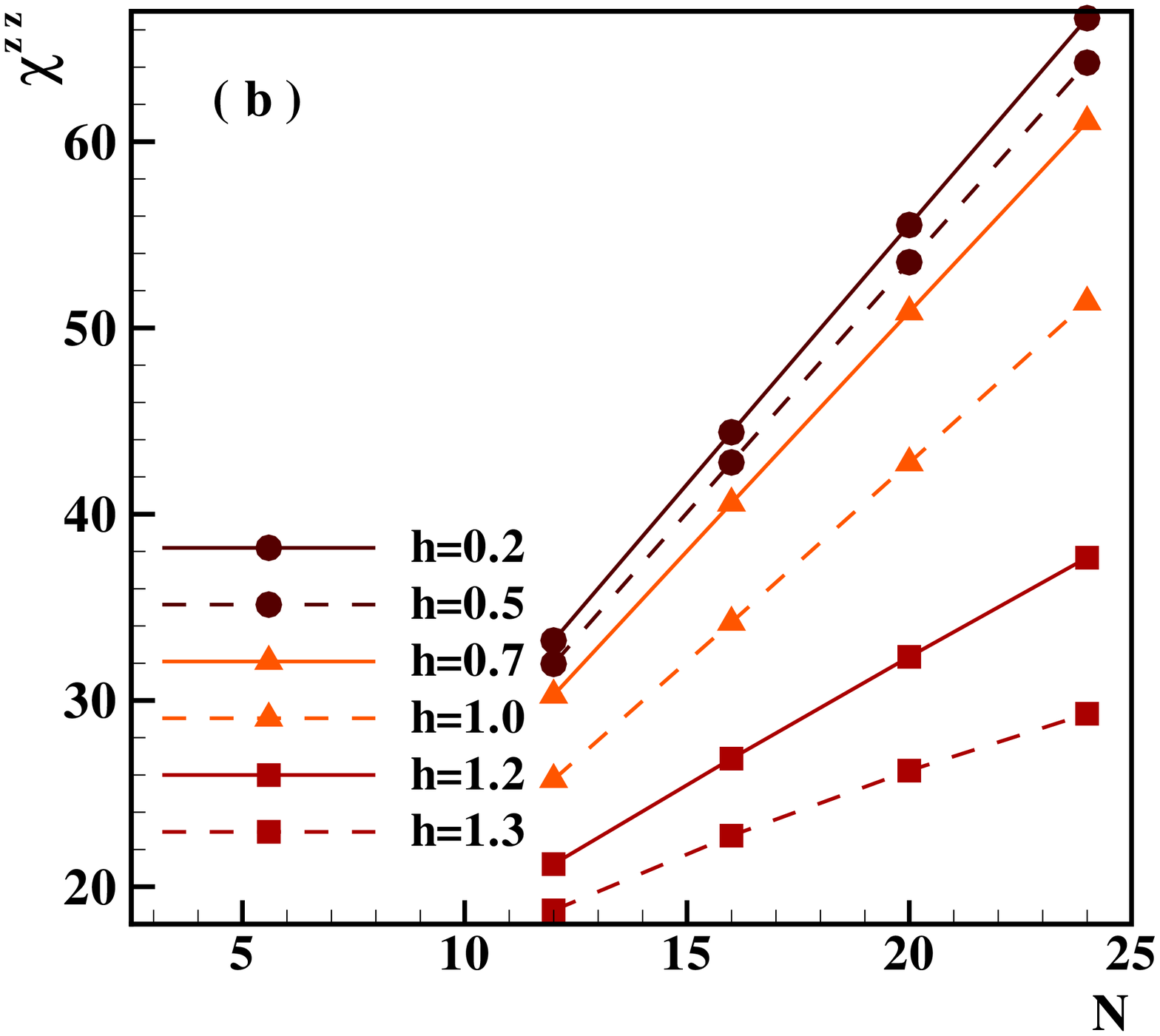}
} \caption{(Color online.) (a) The correlation function of the
stripe antiferromagnetic order parameter $\chi^{zz}/(N/2-1)$ as a
function of the transverse field $h$ for different chain lengths
    $N=12, 16, 20, 24$ and exchanges $L_{1}=1.0$, $J_1=-3.0, J_2=0.5$. (b) The correlation function $\chi^{zz}$
    versus the chain length $N$ for different values of the field
and exchanges $L_{1}=1.0$, $J_1=-3.0, J_2=0.5$. } \label{chi-zz}
\end{figure*}
The final insight into the nature of different phases be related
to the effect of the field in the region (III.) Which can be
obtained by studying the correlation function $\chi^{zz}$. In
Fig.\ref{chi-zz}(a), we have plotted $\chi^{zz}/(N/2-1)$ as a
function of the field $h$ for different values of the chain length
$N=12, 16, 20, 24$ and exchanges $L=1.0$, $J_1=3.0$, $J_2=-0.5$.
It is completely clear, that there is stripe-antiferromagnetic
ordering along the $z$ axis in the region $h<h_c=1.2\pm0.1$. The
induced quantum fluctuations by increasing $h$ from zero,
decreases the stripe-antiferromagnetic order from almost
saturation value.
 Again, to check the existence of the stripe-antiferromagnetic order in the
thermodynamic limit $N \longrightarrow \infty$ of the system in
the region $h<h_c=1.2\pm0.1$, we have plotted in
Fig.\ref{chi-zz}(b) the $N$ dependence of $\chi^{zz}$ for
different values of field $h<h_c=1.2\pm0.1$. As is seen from this
figure in the region $h<h_c=1.2\pm0.1$ there is a divergent
behavior which shows that the stripe-antiferromagnetic order is
true long-range order. By investigating the $N$ dependence of
$\chi^{zz}$ for different values of $J_1/L_1$, we found that there
is no long-range stripe-antiferromagnetic order for fields larger
than the $h_c=1.2\pm0.1$.


\section{conclusion}\label{sec-III }

In this paper we considered the 1D quantum compass model (QCC) and
built a clear picture of different magnetic phases induced by an
external transverse magnetic field at zero temperature.  This
picture is made step by step. At the first step, using the
analytical spinless fermion approach we found that, depending on
the values of the couplings, by increasing the transverse magnetic
field,  one or two quantum phase transitions occur in the ground
state magnetic phase diagram of the QCC. These quantum phase
transitions belong to the universality class of the
commensurate-incommensurate phase transition.

In the second step, we did an accurate  numerical experiment using
the Lanczos method. This part of our work, helped us to increasing
the brightness of the ground state phase diagram picture. We have
implemented the Lanczos method to numerically diagonalize finite
chains. Using the exact diagonalization results, first we have
calculated the energy gap. In complete agreement with our
analytical results, we showed that depending on the values of the
couplings, by increasing the transverse magnetic field, the energy
gap closed at one or two critical fields. Then we also studied the
magnetization process in all regions of the phase diagram. By
comparing the magnetization curves, we showed that the suggested
zero-field hidden ordered phases have different behaviors in
presence of a transverse field.  The hidden order phase in the
region (I.), will not surrender versus the TF, but the hidden
order phase in the region (IV.) as soon as the TF applied will
surrender versus field and spins start to aline in the field
direction. Finally, to distinguish the kind of magnetic orders we
plotted the spin-spin correlation functions and found that the
spin-flop long-range order can be induced in a sector of the
ground state magnetic phase diagram.

\section{acknowledgments}

Authors would like to thank, J. H. H. Perk, H. Johannesson, E. Eriksson, and J. Abouie for useful comments and interesting discussions. MM would like to thank from  A. Davody and  H. R. Afshar for comments on mathematical calculations.




\vspace{0.3cm}

\end{document}